\documentclass[print]{revtex4}
\usepackage{amsmath,amssymb}
\usepackage{graphicx}

\begin{document}

\title{\large \bf Simultaneous creations of discrete-variable entangle state and single-photon-added coherent state}

\author{Yan Li,$^{1,2,3}$\footnote{Electronic address: liyan@wipm.ac.cn} Hui Jing,$^{1,2,4}$\footnote{Electronic address: jinghui73@gmail.com}and Ming-Sheng Zhan$^{1,2}$\footnote{Electronic address: mszhan@wipm.ac.cn}}

\affiliation{$^{1}$State Key Laboratory of Magnetic Resonance and
Atomic and Molecular Physics,\\
 Wuhan Institute of Physics and Mathematics, Chinese Academy of Sciences, Wuhan 430071, P. R. China\\
 $^{2}$Center for Cold Atom Physics, Chinese Academy of Sciences, Wuhan
 430071, P. R. China\\
$^{3}$Graduate School of the Chinese Academy of Sciences, Beijing
100080, P. R. China\\
$^{4}$Department of Physics, The University
of Arizona, 1118 East 4th Street, Tucson, AZ 85721}

\begin{abstract}
The single-photon-added coherent state (SPACS), as an intermediate
classical-to-purely-quantum state, was first realized recently by
Zavatta \emph{$et~al.$} (Science 306, 660 (2004)). We show here
that the success probability of their SPACS generation can be
enhanced by a simple method which leads to simultaneous creations
of a discrete-variable entangled state and a SPACS or even a
hybrid-variable entangled SPACS in two different channels. The
impacts of the input thermal noise are also analyzed. \\

\end{abstract}

\baselineskip=16pt

\maketitle

\noindent The preparations of a nonclassical quantum state are
essential in current quantum information science. Many schemes
have been formulated in the past decade based on the nonlinear
mediums or the technique of conditional measurements
\cite{1,2,3,4,5,6,7,8}. For example, Sanders proposed a concept of
entangled coherent state (ECS) by using a nonlinear interferometer
\cite{6}. Dakna \emph{$et~al.$} used a conditional measurement on
the beam-splitters (BS) to create several kinds of nonclassical
states \cite{7}. Agarwal and Tara presented a hybrid nonclassical
state called a photon-added coherent state (PACS) which exhibits
an intermediate property between a classical coherent state (CS)
and a purely quantum Fock state (FS) \cite{8}. Recently, by using
a type-I beta-barium borate(BBO)crystal, a single-photon detector
(SPD) and a balanced homodyne detector, Zavatta \emph{$et~al.$}
experimentally created a single-photon-added coherent state
(SPACS) which allowed them to first visualize the
classical-to-quantum transition process \cite{9}. By applying the
Sanders ECS, a feasible scheme was proposed to create even an
entangled SPACS (ESPACS) from which one achieves a type-II hybrid
entanglement of the quantum FS and the classical CS [5]. Then it
is clear that, for the purpose of practical applications, the rare
success probability of the SPACS generation in the experiment of
Zavatta \emph{$et~al.$} \cite{9} should be largely improved. In
this paper, by directly combining many parametric amplifiers in
the original scheme of Zavatta \emph{$et~al.$} \cite{9}, a simple
but efficient method is presented to significantly improve the
success probability of their SPACS generation, which is made
possible by simultaneously preparing a discrete-variable entangled
$W$ state \cite{10} and a SPACS (hybrid-variable quantum state) in
two different channels.
 \noindent
\begin{figure}[ht]
 \setlength{\unitlength}{1.0mm}
             \centering
\includegraphics[width=0.6\columnwidth]{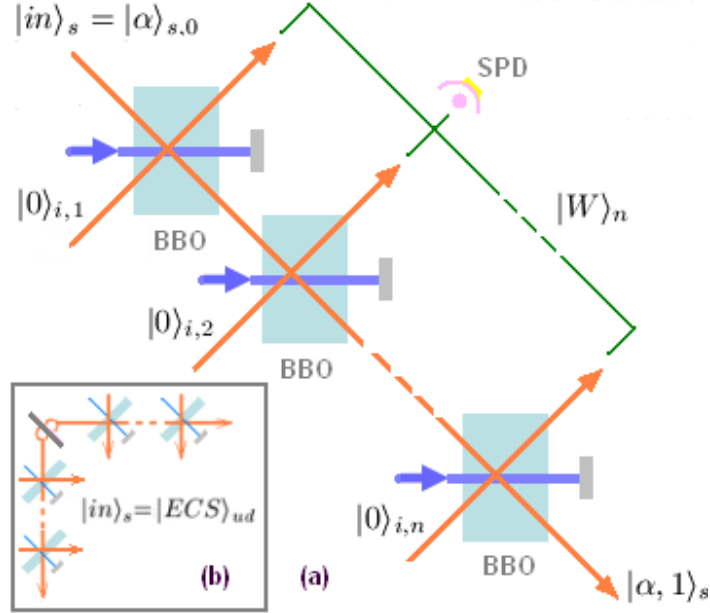}
\caption{(Color online) Schematic diagram to simultaneously create
the discrete-variable entangled $W$ state and the SPACS in two
different channels. The input signal is in (a) a classical CS
\cite{9} or (b) an ECS \cite{5,6}.}
\end{figure}

\indent The PACS $|\alpha,m\rangle$, firstly introduced by Agarwal
and Tara \cite{8}, is defined as
\begin{equation}\label{eqn:1}
|\alpha,m\rangle=\frac{\hat{a}^{\dag
m}|\alpha\rangle}{[m!L_{m}(-|\alpha|^{2})]^{1/2}},
\end{equation}
where $\hat{a}$ $(\hat{a}^{\dag})$ is the photon annihilation
(creation) operator and $L_{m}(x)$ is the Laguerre polynomial ($m$
is an integer),
$L_{m}(x)=\sum_{n=0}^{m}\frac{(-1)^{n}x^{n}m!}{(n!)^{2}(m-n)!}$.
Obviously, for $\alpha$ (or $m$) $\rightarrow0$,
$|\alpha,m\rangle$ reduces to the FS (CS). The interesting
properties of PACS, as an intermediate classical-to-quantum state,
was studied in Ref. \cite{8}. We note that another different
intermediate state called the displaced FS:
$|DFS\rangle=D(\alpha)|n\rangle=\exp(\hat{a}^{\dag}\alpha-\hat{a}\alpha^{*})|n\rangle$,
is obtained by acting upon the FS with the displacement operators
\cite{4}. The PACS is, however, obtained via the successive
one-photon excitations on a classical CS light.

\indent Now we consider the parametric down-conversion process
(type-I BBO crystal) in which one photon incident on the
dielectric with $\chi^{2}$ nonlinearity breaks up into two new
photons of lower frequencies. In the steady state, we always have
$\omega_{p}=\omega_{s}+\omega_{i}$, where $\omega_{p}$ is the pump
frequency, and $\omega_{s}$ or $\omega_{i}$ is the signal or idler
frequency. Under the phase matching condition, the wave vectors of
the pump, signal and idler photons are related by
$\vec{k}_{p}=\vec{k}_{s}+\vec{k}_{i}$ (momentum conservation). The
signal and idler photons appear simultaneously within the
resolving time of the detectors and the associated electronics
\cite{11}. The Hamiltonian in the interaction picture is then: $
\hat{H}_I=\hbar
g[\hat{a}_{s}^{\dag}\hat{a}_{i}^{\dag}\hat{a}_{p}+\hat{a}_{s}\hat{a}_{i}\hat{a}_{p}^{\dag}],
$ where the real coupling constant $g$ contains the nonlinear
susceptibility $\chi^{2}$. The free-motion parts of the total
Hamiltonian commute with $\hat{H}_I$. Thereby one pump photon is
converted into one signal and one idler photon.

\indent We consider for simplicity the signal and idler lights
with a same polarization and the well-defined directions, and an
incident intense pump for which the quantum operator $\hat{a}_{0}$
can be treated classically as $\hat{a}_{0}\rightarrow iV$. Then
for an input CS:
$|\psi(0)\rangle_1=|\alpha\rangle_{s}|0\rangle_{i}$, the output
state after an interaction time $t$ with one nonlinear crystal
evolves into: $
|\psi(t)\rangle_1=\exp[-i\hat{H}_{I}t/\hbar]|\psi(0)\rangle_1
=\exp[\lambda(\hat{a}_{s}^{\dag}\hat{a}_{i}^{\dag}-\hat{a}_{s}\hat{a}_{i})]|\psi(0)\rangle_1,
$ with an effective interaction time $\lambda=Vgt$. For short
times $t$ compared with the average time interval between the
successive down-conversions, by expanding the exponential we have
($\lambda\ll 1$)
\begin{equation}\label{eqn:1}
|\psi(t)\rangle_1
\approx|\alpha\rangle_{s}|0\rangle_{i}+\lambda|\alpha,1\rangle_{s}|1\rangle_{i}+
\frac{\lambda^{2}}{2}(\hat{a}_{s}^{\dag}\hat{a}_{i}^{\dag}\hat{a}_{s}^{\dag}\hat{a}_{i}^{\dag}-
\hat{a}_{s}\hat{a}_{i}\hat{a}_{s}^{\dag}\hat{a}_{i}^{\dag})|\alpha\rangle_{s}|0\rangle_{i}.
\end{equation}
Due to $\lambda\ll 1$, we can select the first two terms as the
output state, i.e., $
|\psi(t)\rangle_1\approx|\alpha\rangle_{s}|0\rangle_{i}+\lambda|\alpha,1\rangle_{s}|1\rangle_{i}
$. Thus the signal channel mostly contains the input CS, except
for the few cases when a single photon is detected in the output
idler. These $rare$ events stimulate one-photon emission into the
CS and then generate a SPACS in the output signal with a success
probability being proportional to
$p_1^1=|\lambda|^{2}(1+|\alpha|^{2})$. As in the experiment of
Zavatta \emph{$et~al.$} \cite{9}, the one-photon excitation is
selected to avoid the higher-order ones which cannot be
discriminated by the SPD with finite efficiency.

\indent As shown in Fig. 1(a), we consider the direct combination
of a series of two or more (say, $N$) identical optical parametric
amplifiers by assuming for simplicity the same low-gain
($g_{j}=g$, $j=1,2,...N$) and the same classical pumps. For an
input CS in the signal and a vacuum in all the idlers, i.e.,
\begin{equation}\label{eqn:1}
|\psi(0)\rangle_N=|\alpha\rangle_{s,0}|0\rangle_{i,1}|0\rangle_{i,2}\ldots|0\rangle_{i,N},
\end{equation}
the output state is a $(N+1)$-body hybrid entangled state which
reads:
\begin{equation}\label{eqn:1}
|\psi(t)\rangle_N =\prod_{j=1}^N
\exp[-i\hat{H}_{jI}t/\hbar]|\psi(0)\rangle_N,
\end{equation}
where $ \hat{H}_{jI}=i\hbar g
V[\hat{a}_{s,j}^{\dag}\hat{a}_{i,j}^{\dag}-\hat{a}_{s,j}\hat{a}_{i,j}]
$. Based on the conditional detections of SPDs in the idlers, the
PACS $|\alpha,m\rangle$ is generated with the success probability
being proportional to
$p_N^m=N|\lambda^{m}|^{2}m!L_{m}(-|\alpha|^{2})$ ($m\leq N$).
Clearly, for $m=1$ (SPACS), we have a simple relation:
$p_N^1=Np_1^1$, which means that, at least under the ideal
conditions, the success probability of the optical SPACS
generation can be improved $N$ times in comparison with the
original scheme of Zavatta \emph{$et~al.$} \cite{9}.

This significant improvement of the SPACS generation has a simple
physical explanation, i.e., the formation of the $N$-qubit $W$
entangled state in the idlers. In fact, it is easy to verify that,
when a SPACS is achieved in the output signal, one has the
following state in the idlers
\begin{equation}\label{eqn:1}
|\psi(t)\rangle_{N}^{W}=\frac{1}{\sqrt{N}}(|1\underbrace{00\cdots0}_{N-1}\rangle+|01\underbrace{00\cdots0}_{N-2}\rangle+\cdots+|\underbrace{00\cdots0}_{N-1}1\rangle).
\end{equation}
This means that, as long as anyone of the $N$ detectors captures
one photon, a SPACS can be achieved! In other words, an
observation of the SPACS cannot tell us which one of the $N$
detectors was hit by a single photon. It is this quantum
$indistinguishability$ which leads to the simultaneous (and
enhanced) generation of the discrete-variable entangle $W$ state
and the SPACS in two different channels.

Example-1: $N=2$. For an input state:
$|\psi(0)\rangle_2=|\alpha\rangle_{s,0}|0\rangle_{i,1}|0\rangle_{i,2}$,
the output state of the system is
\begin{equation}
|\psi(t)\rangle_2 \approx
|\alpha\rangle_{s}|0\rangle_{i,1}|0\rangle_{i,2}+\lambda|\alpha,1\rangle_{s}
|EPR\rangle_2+
\lambda^2|\alpha,2\rangle_{s}|1\rangle_{i,1}|1\rangle_{i,2},
\end{equation}
where we denote $|EPR\rangle_2=|1\rangle_{i,1}|0\rangle_{i,2}
+|0\rangle_{i,1}|1\rangle_{i,2}$. Obviously, due to $\lambda\ll
1$, we can select the first two terms as the output state and the
SPACS is created in the output signal with a success probability
being proportional to:
$p_2^1=2p_1^1=2|\lambda|^{2}(1+|\alpha|^{2})$. In short, via the
conditional SPDs technique of Zavatta \emph{$et~al.$} \cite{9},
the probabilistic SPACS generation is enhanced $2$ times by
simultaneously creating an EPR-type maximally entangled state and
the SPACS in two different channels.

Example-2: $N=3$. Still for a classical CS input signal, we
achieve the output state of the system:
\begin{equation}
|\psi(t)\rangle_3 \approx
|\alpha\rangle_{s}|0\rangle_{i,1}|0\rangle_{i,2}|0\rangle_{i,3}+\lambda|\alpha,1\rangle_{s}
|W\rangle_3 +\lambda^2|\alpha,2\rangle_{s} |II\rangle_3
+\lambda^3|\alpha,3\rangle_{s}|1\rangle_{i,1}|1\rangle_{i,2}|1\rangle_{i,3},
\end{equation}
with the $3$-body entangled states
$|W\rangle_3=|1\rangle_{i,1}|0\rangle_{i,2}|0\rangle_{i,3}+|0\rangle_{i,1}|1\rangle_{i,2}|0\rangle_{i,3}
+|0\rangle_{i,1}|0\rangle_{i,2}|1\rangle_{i,3}$ and
$|II\rangle_3=|1\rangle_{i,1}|1\rangle_{i,2}|0\rangle_{i,3}+
|0\rangle_{i,1}|1\rangle_{i,2}|1\rangle_{i,3}+|1\rangle_{i,1}|0\rangle_{i,2}|1\rangle_{i,3}$.
Obviously, due to $\lambda\ll 1$, we still can select the first
two terms as the output state and the SPACS is created with a
success probability being proportional to:
$p_3^1=3p_1^1=3|\lambda|^{2}(1+|\alpha|^{2})$, which is made
possible by simultaneously creating the $3$-body $W$-type
maximally entangled state and the SPACS in two different channels
(see also the Eq. (5)).

As a practical realization, the $N$ idlers can be connected by a
multi-port optical fiber to one SPD since in all of the $N$ idlers
there is maximally one photon to be detected for $\lambda\ll 1$.
We note that, although it is difficult to achieve a large $N$ with
the condition of a strong strength for all the pumps if one uses
some beam-splitting technique on one strong pump laser beam
\cite{13}, this method still can be useful due to the extreme
difficulty to achieve a large nonlinear susceptibility in the
ordinary nonlinear optics mediums \cite{1}. To further improve the
success probability of the SPACS generation, one can even
consider, e.g., the complex technique of electromagnetically
induced transparency (EIT) to get a giant enhancement of the
nonlinear susceptibility in an ultra-cold three-level atomic cloud
\cite{13}.

This repeated BBO method also can be applied to some more
complicated scheme, say, two output signals with an input ECS (see
Ref. [5] or Fig. 1(b)). We consider the concrete example of two
BBO in the upper channel and still one BBO in the down channel.
The initial state with an input two-body ECS signal can be written
as
$|\psi\rangle_{in}=|ECS\rangle_{ud}|0\rangle_{ui,1}|0\rangle_{ui,2}|0\rangle_{di}$,
where the Sanders ECS is \cite{6}
\begin{equation}
|ECS\rangle_{ud}=\frac{1}{\sqrt{2}}[e^{-i\pi/4}|i\beta\rangle_{us}|i\alpha\rangle_{ds}
+e^{i\pi/4}|-\alpha\rangle_{us}|\beta\rangle_{ds}].
\end{equation}
Then the following output state is achieved as (up to first order
of $\lambda$)
\begin{equation}
|\psi\rangle_{out}=|\psi\rangle_{in}
+\lambda|ESPACS\rangle_{ud}^I|EPR\rangle_{u,i}|0\rangle_{di}
+\lambda|ESPACS\rangle_{ud}^{II}|0\rangle_{ui,1}|0\rangle_{ui,2}|1\rangle_{di},
\end{equation}
where the $hybrid$ entangled states or the ESPACS \cite{5} are
\begin{eqnarray}
|ESPACS\rangle_{ud}^I=\frac{1}{\sqrt{2}}[e^{-i\pi/4}|i\beta,1\rangle_{us}|i\alpha\rangle_{ds}
+ e^{i\pi/4}|-\alpha,1\rangle_{us}|\beta\rangle_{ds}],\nonumber \\
|ESPACS\rangle_{ud}^{II}=\frac{1}{\sqrt{2}}[e^{-i\pi/4}|i\beta\rangle_{us}
|i\alpha,1\rangle_{ds} +
e^{i\pi/4}|-\alpha\rangle_{us}|\beta,1\rangle_{ds}],
\end{eqnarray}
and
$|EPR\rangle_{u,i}=|1\rangle_{ui,1}|0\rangle_{ui,2}+|0\rangle_{ui,1}|1\rangle_{ui,2}$.
This clearly shows that we can simultaneously create the two-body
EPR-type entangled state and the ESPACS in two different channels
(i.e., the two idlers or the two signals), which also makes the
success probability of achieving the SPACS in the upper signal be
$2$ times than in the down signal. The similar results can be
obtained easily for some more general configurations, such as the
simultaneous creations of the $W$ entangled state and the ESPACS.

Finally we give a simple analysis about the impacts of possible
input thermal noise on the SPACS generation in the experiment of
Zavatta \emph{$et~al.$} \cite{9}. With a perfect vacuum for all
the input signals, we only need to consider some mixed thermal
noise in the input CS signal. The finite temperature effect can be
described by the Takahashi-Umezawa formalism of thermo-field
dynamics (TFD) in which the thermo vacuum state is defined as
\cite{14}: $|0\rangle_T\equiv \hat{H}(\theta)|0 \tilde{0}\rangle$,
where the new vacuum state $|0\tilde{0}\rangle$ belongs to the
double Hilbert space determined by the tilde conjugate, and the
heating operator:
$\hat{H}(\theta)=\exp[-\theta(\hat{a}_s\hat{\tilde{a}}_s-\hat{a}_s^\dagger\hat{\tilde{a}}_s^\dagger)]$
provides a $thermal$ Bogoliubov transformation:
\begin{equation}
\hat{H}^\dagger(\theta)\hat{a}_s\hat{H}(\theta)\equiv\hat{b}_s=u(\beta)\hat{a}_s+v(\beta)\hat{\tilde{a}}_s^\dagger,
\end{equation}
where $u(\beta)=\cosh\theta$, $v(\beta)=\sinh\theta$ and the new
quasi-particle operators also satisfy
$[\hat{b}_s,\hat{b}_s^\dagger]=1$. The photons of the thermal
vacuum obey the normal Bose-Einstein distribution, i.e.,
$\bar{n}\equiv \langle 0 \tilde{0}|\hat{b}_s^\dagger \hat{b}_s|0
\tilde{0}\rangle=\sinh^2{\theta}= (e^{\beta\omega}-1)^{-1}$ with
$\beta=(k_B T)^{-1}$, from which we have:
$u(\beta)=\sqrt{\bar{n}+1}$ and $v(\beta)=\sqrt{\bar{n}}$. It is
this expression which determines the heating coefficient $\theta$
in the heating operator $\hat{H}(\theta)$. Formally, the
time-evolution operator of the parametric amplifier is also some
"heating" operator but with $\theta\rightarrow\lambda$. Using the
TFD formalism, we can take the initial state of the low
temperature system ($v(\beta)\ll u(\beta)$) as the mixed
coherent-thermal fields \cite{14}:
$|in\rangle_s=\hat{H}(\theta)\hat{D}(\alpha)|0\tilde{0}\rangle_s=\hat{H}(\theta)|\alpha\tilde{0}\rangle_s$,
and an additional fictitious displacement operator can also be
introduced for a high temperature. Therefore the output state of
the system is simply written as (up to the first order of
$\lambda$)
\begin{equation}
|\psi(t)\rangle_1=\hat{H}(\theta)
\exp[\lambda(\hat{b}_{s}^{\dag}\hat{a}_{i}^{\dag}-\hat{b}_{s}\hat{a}_{i})]|\alpha\tilde{0}\rangle_s|0\rangle_i
\approx|in\rangle_s|0\rangle_i + \lambda
u(\beta)\hat{H}(\theta)|\alpha\tilde{0},1\rangle_s|1\rangle_i.
\end{equation}
This, by ignoring the unchanged fictitious mode, leads to a
conversion: $\hat{H}(\theta)|\alpha\rangle_s|0\rangle_i\rightarrow
\hat{H}(\theta)|\alpha,1\rangle_s|1\rangle_i$ with the success
probability of $p_1^1(\beta)=\lambda u(\beta)$, which means that,
even with the input mixed coherent-thermal fields, the SPACS still
$can$ be achieved with an "amplified" success probability, i.e.,
$\lambda\rightarrow\lambda u(\beta)$. However, for an input
thermalized CS state instead of an ideal CS state, the achieved
SPACS in fact is also a thermalized SPACS instead of an ideal
SPACS. Thereby it is not surprising to expect that the quantum
statistical properties, including the Wigner functions observed in
the experiment of Zavatta \emph{$et~al.$} \cite{9}, can experience
some deformation tending to weaken or smear its nonclassical
features. The similar results is readily obtained for the repeated
amplifiers case. The SPACS generation scheme of Zavatta
\emph{$et~al.$} \cite{9} is thus confirmed to be robust to some
thermal noise in the input CS signal.

In conclusion, we propose a simple method to simultaneously create
the discrete-variable entangled state and the SPACS or even the
hybrid-variable entangled SPACS (ESPACS) in two different
channels. It is interesting to observe that the formation of
quantum entanglement or indistinguishability can lead to the
improvement of the SPACS generation. Many other techniques to
improve the SPACS generations may exist such as a high-frequency
time-resolved balanced homodyne detection and a mode-locked laser
(see Zavatta \emph{$et~al.$} \cite{9}), our simple method of
repeated BBO here also can be of some values, taking into account
of the important applications of both the SPACS and the entangled
$W$ state. Although the $W$ state can be generated by other more
efficient ways, this is the first time to simultaneously create
both the PACS and the $W$ state. Another interesting point of this
proposal could be the possibility to select the two-photons exited
coherent states simply by detecting coincidences of SPDs placed at
the two idler outputs (see Eq. (6)) \cite{15}.

From the experimental point of view using more than one crystal is
feasible but much more complicate. There are always some realistic
problems such as the imperfect elements which affects the
generation efficiency and the fidelity of the desired output
state, and many authors analyzed in detail such losses as well as
the suggestions of improving the efficiency and fidelity \cite{16,
17}. The practical limitations of the detector can be a main
difficult problem for the SPACS generation and one should use a
SPD bearing a lower dark count rate and shorter resolution time on
the premise of same efficiency. The input CS light intensity
should be lowered to get a higher fidelity \cite{16}. Due to the
finite crystal size and the spatial location of the idler-signal
output photons, some narrow spatial and frequency filters should
be placed in the idler output before the detector. Of course,
since there are many other lossy factors like the
environment-induced damping associated with the repeated system
\cite{17}, the enhancement in the production rate may not be so
high to justify the increasing complexity of the setup. This
difficulty arises also from the fact that the generation of
multi-qubit entanglement is a difficult task in the present
experiments \cite{18}. However, since our simple method can
generate the PACS with $m>1$ with a higher production rate with
respect to the single crystal case and it requires the single
photon detectors (SPDs) only [15], it can be an interesting and
challenging scheme for the future experiment.
\bigskip

\noindent {\it Note Added.} After finishing this work, we found a
formally similar but different idea of using repeated PDC for
state control by Prof. A. Lvovsky group, see
http://qis.ucalgary.ca/quantech/repeated.html.

\bigskip

\noindent H. J. is grateful to P. Meystre and A. Zavatta for their
kind help and very useful discussions. This work was supported
partially by NSFC (10304020) and Wuhan Sunshine Program (CL05082).

%\section{}

%\section{Results}
%\section{Conclusions}
%\indent

%%%%%%%%%%%%%%%%%%%%%%%%%%%%%%%%%%%%%%%%%%%%%

%%%%%%%%%%%%%%%%%%%%%%%%%%%%%%%%%%%%%%%%%%%%%

%\bigskip

\end{document}